\documentstyle[12pt]{article}

\begin{document}
\vspace*{.5cm}

\begin{center}
{\LARGE
Long-term Relaxation of a Composite System in Partial Contact with a Heat Bath
\\}
\vspace{.8cm}
{\large Naoko Nakagawa, Kunihiko Kaneko $^1$ and Teruhisa S. Komatsu $^2$}\\
\end{center}

\vspace{0.1cm}
{
Department of Mathematical Sciences, 
Ibaraki University,
Mito, Ibaraki 310-8512\\
$^1$ Department of Pure and Applied Sciences,
College of Arts and Sciences,
University of Tokyo, Tokyo 153-8902\\
$^2$ Department of Environmental Science,
JAERI, Tokai, Naka, Ibaraki 319-1195
%Japan Atomic Energy Research Institute, Tokai, Naka, Ibaraki 319-1195
}

\vspace{0.4cm}

\begin{abstract}
We study relaxational behavior from a highly excited state for 
a composite system in partial contact with a heat bath,
motivated by an experimental report of
long-term energy storage in protein molecules.
The system consists of two coupled elements: 
The first element is in direct contact with a heat bath,
while the second element interacts only with the first element.
Due to this indirect contact with the heat bath, energy injected into
the second element dissipates very slowly, according to a power law, 
whereas that injected into the first one exhibits exponential dissipation.
The relaxation equation describing this dissipation is obtained analytically 
for both the underdamped and overdamped limits.
Numerical confirmation is given for both cases.
\end{abstract}

%\kword{long-term energy storage, indirect contact with a heat bath, 
%Boltzmann-Jeans conjecture, protein}

\section{Introduction}

When a system is in contact with a heat bath,
in the most common situation it relaxes toward the equilibrium state 
rather rapidly,
typically in a manner characterized by exponential decay, except at a
phase transition point.
For instance, a hot particle in cold water generally loses its energy 
very quickly.
However, does such rapid relaxation of a high-energy state hold generally?
Consider, for example, a system composed of
several elements that is in contact with a heat bath only
through its outer part.  A protein molecule in water
represents such an example, since the hydrophobic part of a protein molecule 
will be segregated from an aqueous solution \cite{note}.
In such a situation, the relaxation process of the internal part naturally
depends on the dynamics of the whole system, including the temporal and spatial
relationship between the internal and external parts.
Indeed, it has been reported that proteins excited into high-energy states, 
exhibit very slow relaxation,
compared with that seen in standard dissipation processes
in systems of similar size
\cite{Ishijima_Yanagida}.
Such long-term energy storage may be necessary for proteins to work 
functionary.
For instance, consider a thermal ratchet model 
\cite{Vale_Oosawa, Magnasco, Julicher_Prost}
for motor proteins that work in a heat bath.
In the model, the existence of both high and low temperature domains 
in proteins are assumed, to generate a directive motion.
Then, for such a mechanism to work, it should be answered how 
the proteins maintain a high energy (i.e., keep an effectively 
high temperature state) at a localized domain within them.

In connection with this long-term energy storage
in proteins, two of the present authors \cite{Nakagawa_Kaneko} have
studied a model of a coupled pendulum in partial contact with a heat bath.
It was found that when a pendulum that is not in contact with a heat bath
 is highly excited, the energy dissipates only slowly, approximately with 
logarithmic decay.
This slow relaxation seems to be consistent with the Boltzmann-Jeans conjecture
\cite{Jeans, Galgani_Benettin}
for very slow relaxation in Hamiltonian systems without contact to a heat bath.
In such systems, the relaxation time increases exponentially of 
some positive power of the injected energy 
\cite{Baldan_Benettin,Nakagawa_Kaneko}.

When all elements in a Hamiltonian system are in full contact with a heat bath,
the mechanism allowing for slow relaxation may be destroyed completely.  
On the other hand, when a system contacts
a heat bath only through an external part,
the internal part can evolve according to its `own' dynamics.
For this reason, the internal Hamiltonian dynamics
may have a strong influence on the relaxation process, and this can lead
to a slow relaxation process and long-term energy storage.
It is thus important to study how a system in partial contact
with a heat bath insulates its internal part from the effect of the heat bath.

To study such behavior of Hamiltonian systems in partial contact with a
 heat bath, we adopt a simple model, introduced in \S 2, consisting of
two coupled elements:
The first element is in direct contacted with a heat bath
to allow for the dissipation of the energy,
while the second element interacts only with the first element,
and not directly with the heat bath.
We study relaxation from a far-from-equilibrium state
by preparing the second element in a highly excited state with
 a large kinetic energy.
The first element, on the other hand, remains always in
near-equilibrium with the heat bath.
It should be noted that, without the heat bath, the presently studied 
Hamiltonian system is always integrable and does not relax,
even when prepared in a highly excited state.
Therefore the relaxation of the second element to thermal equilibrium is
due to the indirect effect of the heat bath.
In \S 3, the relaxation equation for the underdamped limit is obtained,
while the case of the overdamped limit is studied in \S 4.
The validity of these equations is confirmed by numerical simulations 
discussed in \S 5.
Relaxation in the intermediate damping case is examined in \S 6.
The results of numerical experiments for this case are also reported there.
{}From these analytical and computational studies,
two kinds of power law are obtained for the underdamped and overdamped cases,
respectively. It is found that these power laws are independent 
of the temperature.

\section{Model System}

To study relaxation of a system with internal degrees of freedom that
is in partial contact with a heat bath,
we consider a Hamiltonian system with $2$ degrees of freedom
of the general form
\begin{equation}
H=\frac{p_1^2}{2}+\frac{p_2^2}{2}+U(\theta_1-\theta_2),
\label{eqn:originalH}
\end{equation}
where $U(\theta)$ is a periodic function 
of period $2\pi$ satisfying $|U|<\infty$.
Considering the system to be in partial contact with a heat bath,
and adopting the standard representation of the heat bath, we take the 
motion of the elements to be given by the following Langevin equations:
\begin{equation}
\left\{
\begin{array}{rcl}
\dot p_1 &=& - \displaystyle\frac{\partial U}{\partial \theta_1}
- \gamma p_1+\xi(t), \\\\
\dot p_2 &=& - \displaystyle\frac{\partial U}{\partial \theta_2}.
\end{array}
\right.
\label{eqn:Langevin}
\end{equation}
Here the random force $\xi(t)$ due to the heat bath consists of 
a Gaussian white noise satisfying $\langle \xi(t)\rangle =0$
and $\langle \xi(t_1)\xi(t_2) \rangle
={2\gamma T}\delta(t_2-t_1)$, with $\langle \cdot \rangle$
representing the temporal average. (We use units in which $k_B=1$).
The parameter $\gamma$ is the dissipation coefficient characterizing 
the dissipation to the heat bath,
and $T$ is the temperature of the heat bath.
With such contact to a heat bath, it is generally expected that a system of
this kind will relax to equilibrium.
The change of the total energy $E$ due to dissipation to the heat bath
can be written 
\begin{equation}
\dot E = -\gamma p_1^2+p_1\xi(t),
\label{eqn:relax-E}
\end{equation}
and we see that $E$ remains constant if there is no contact ($\gamma=\xi=0$) 
with the heat bath.

Using the set of variables
$\Theta \equiv \theta_1+\theta_2, \phi \equiv \theta_1-\theta_2$
and $P \equiv p_1+p_2=\dot\Theta$,
we rewrite the Hamiltonian as
\begin{equation}
H=\frac{P^2}{4}+\frac{\dot\phi^2}{4}+U(\phi).
\label{eqn:Hamiltonian}
\end{equation}
When there is no heat bath (i.e., for $\gamma=\xi=0$), 
the equation of motion is completely separable in these variables:
$\dot{P}=0$, $\ddot{\phi}=-2{\partial U}/{\partial \phi}$.
Thus we see that in this case this system is completely integrable, and 
the total momentum $P$ is a conserved quantity.

In the following sections, we study this system analytically 
to elucidate the relaxational behavior in certain limits.
The results of numerical simulations of some models are given in \S 5 and \S 6.

\section{Underdamped Limit}

In this section we consider (2) in the underdamped limit, 
i.e. with a small value of $\gamma$.
With the presence of this weak damping term,
the total energy $E$ and total momentum $P$, which are conserved quantities 
for $\gamma=0$, become slow variables.

\subsection{Derivation of relaxation equation}

Let us consider the situation in which the system is initially far from 
equilibrium, with very large  $|p_2| \gg |p_1|$.  The rotation of
$\phi$, accordingly, is much faster than its change due to dissipation.
Then, the terms $\partial U/\partial\theta_1$ and $\partial U/\partial\theta_2$
in (\ref{eqn:Langevin}) change so fast that their effects 
on the dynamics of $p_1$ and $p_2$ are averaged out over a time scale
 relevant for these slower variables.
Here we have to distinguish the orders of characteristic time scales.
We have the following relations : (the time scale of $\xi(t)$) $\ll$
 (the time scale of rotation $2\pi |\dot\phi|^{-1}$)
$\ll$ (the time scale of relaxation $\gamma^{-1}$).
Now, we introduce a slower time scale $\tau$ satisfying
$2\pi |\dot\phi|^{-1}\ll \tau \ll \gamma^{-1}$,
and consider averages over this time scale,
$\tilde X(t)\equiv\frac{1}{\tau} \int^{t+\tau}_{t} X{\rm d}t$.
Using variables averaged over this time scale, 
the mode of fast rotation is eliminated.

The equations of motion for the slow time scale are obtained by integrating
(\ref{eqn:Langevin}) from $t$ to $t+\tau$. This yields
\begin{equation}
\left\{
\begin{array}{rcl}
\displaystyle{\frac{{\rm d}\tilde p_1}{{\rm d}t}} &=& -\gamma \tilde p_1 -\tilde{U}'
+\tilde\xi,\\\\
\displaystyle{\frac{{\rm d}\tilde p_2}{{\rm d}t}} &=& \tilde{U}',
\end{array}\right.
\label{eqn:Langevin-U}
\end{equation}
where  $U'={\rm d}U/{\rm d}\phi=\partial U/\partial\theta_1=-\partial U/\partial\theta_2$,
$\tilde{U}'\equiv \frac{1}{\tau} \int^{t+\tau}_{t} {\rm d}U(\phi(t'))/{\rm d}\phi\, {\rm d}t'$
and $\langle \tilde\xi(t)\tilde\xi(t')\rangle=2\gamma T \delta(t-t')$,
with $\langle\tilde\xi(t)\rangle=0$. The $\delta$-function here 
is defined with respect to the time scale $\tau$.

In the limit $|\dot\phi| \rightarrow \infty$,
the averaged variables can simply be replaced by
their uniform averages over $\phi$, defined generally by \,
$\overline{X}=\int_0^{2\pi}X(\phi){\rm d}\phi/(2\pi)$. Then we obtain
\begin{equation}
\left\{
\begin{array}{rcl}
\displaystyle{\frac{{\rm d}\overline{p_1}}{{\rm d}t}} &=& 
-\gamma \overline{p_1}+\overline\xi,\\\\
\displaystyle{\frac{{\rm d}\overline{p_2}}{{\rm d}t}} &=& 0,
\end{array}\right.
\end{equation}
where $\overline{U'}=0$ for periodic potentials.
Then, as the lowest-order estimation in this limit,
we have
\begin{equation}
\tilde p_1\simeq \tilde p_1(0)\, {\rm e}^{-\gamma\, t},\qquad
\tilde p_2=\tilde p_2(0),
\label{eqn:decay}
\end{equation}
starting in a state far from equilibrium with large values of $|p_1|$
 and $|p_2|$.
Beginning in such a state, after a time scale of $O(\gamma^{-1})$, 
the first element approaches thermal equilibrium ($p_1^2 \sim T$), while,
within this lowest-order approximation,
the second element continues to satisfy $p_2=p_2(0)$
and does not evolve toward equilibrium.

To observe the relaxation process of the second element, 
we need the next order estimation. 
Since $\dot\phi$ is finite and can depend on $\phi$,
in general the value of a time-averaged variable $\tilde X$ deviates from 
that of the corresponding phase-averaged variable $\overline{X}$.
We now derive the correction to the equations of motion that
this deviation introduces to first order in $|\dot\phi|^{-1}$,
and thereby obtain the relaxational equations
for the second element and the total energy $E$.

After the exponential decay derived by (\ref{eqn:decay})
occurring on a time scale $O(\gamma^{-1})$,
$p_1$ is approximately equal to its equilibrium value while 
$p_2$ is still large. 
Then the relations
\begin{equation}
|p_1|\sim \sqrt{T} \ll p_2\simeq P
\label{eqn:order}
\end{equation}
hold, where $P$ is the total momentum of the system
and is large enough for the fast rotational motion of 
the second element to be maintained.  
From (\ref{eqn:order}), we straightforwardly obtain
$\dot{\phi}\simeq -P$. 
The total energy $E$ of the system can then be expressed by the variables 
$P$, $\phi$ and $p_1$ as
\begin{equation}
E=\frac{P^2}{2}+U(\phi) -p_1 P+p_1^2 ,
\label{eqn:Delta}
\end{equation}
where $U(\phi)$, $p_1 P$ and $p_1^2$ are much smaller than $P^2/2$.

Since $|\dot\phi|$ is sufficiently large and $\gamma$ is sufficiently small, 
$E$ and $P$ are almost conserved over the time scale $\tau$.
To consider the slow evolution of $E$ and $P$,
we eliminate the fast variable $\phi$,
 using the time averaged variables $\tilde X$.
Within the slow time scale, the value of $E$ can then be expressed as
\begin{equation}
E = \frac{P^2}{2}+\tilde U -\tilde p_1\,P 
+ {\tilde p_1}^2,
\label{eqn:meanH}\\
\end{equation}
where $\frac{1}{\tau}\int_t^{t+\tau}p_1^2 {\rm d}t'$ is approximated by
 ${\tilde p_1}^2$ on the time scale $\tau \ll\gamma^{-1}$.
{}From the relation between (\ref{eqn:Hamiltonian}) and (\ref{eqn:meanH}),
the value of $\dot \phi$ is estimated as
\begin{eqnarray}
\dot\phi&=&-\sqrt{4(E-U(\phi))-P^2}=-P\sqrt{1-\frac{4\tilde p_1}{P}
-\frac{4(\delta U-\tilde p_1^2)}{P^2}} \nonumber \\ 
&=& -P+2\tilde p_1 +\frac{2\delta U}{P}
+ \frac{4\tilde p_1\delta U}{P^2}
+ O(\frac{\delta U^2}{P^3},\frac{T\delta U}{P^3}),
\label{eqn:phi}
\end{eqnarray}
where $\delta U\equiv U-\tilde U$.
The last expansion is valid if the conditions 
$|\delta U| \ll P^{2}$ and $T\ll P^2$
are satisfied.
The momenta of the two elements are then given by
\begin{equation}
p_1 = \frac{ P + \dot\phi }{2}
= \tilde p_1 +\frac{\delta U}{P}+\frac{2\tilde p_1\delta U}{P^2} 
+ O(\frac{\delta U^2}{P^3},\frac{T\delta U}{P^3}),
\label{eqn:p1-U}
\end{equation}
This result reveals that both elements oscillate rapidly 
around their averages $\tilde p_1$ and $\tilde p_2$ 
($=P-\tilde p_1$).
We see that the amplitude of oscillation becomes smaller as the 
second element rotates faster (i.e., for larger $p_2$ and $P$).

In order to obtain the relaxation equation for $E$, 
we substitute (\ref{eqn:p1-U}) into (\ref{eqn:relax-E}).
By considering only the case with a sufficiently low temperature, and
neglecting terms of $O(TP^{-2})$, we obtain
\begin{equation}
\dot E=-\frac{\gamma\delta U^2}{P^2}+\frac{2\tilde p_1\delta U}{P}
+\frac{\delta U \xi}{P}+\tilde p_1(-\gamma\tilde p_1 +\xi) 
+O(\frac{\delta U^2}{P^3},\frac{T}{P^2}).
\end{equation}
This equation includes fast variables,
while the total energy $E$ is a slow variable. 
It is thus appropriate to average out the fast change in this equation.
Doing so, we obtain
\begin{equation}
\frac{{\rm d}E}{{\rm d}t}=
-\frac{\gamma}{P^2} \{\frac{1}{\tau}\int_{t}^{t+\tau}\delta U^2{\rm d}t'\}
+\frac{1}{P}\{\frac{1}{\tau}\int_{t}^{t+\tau}\delta U \xi(t'){\rm d}t'\}
+\frac{{\rm d}}{{\rm d}t}(\frac{\tilde p_1^2}{2}+\tilde U) 
+O(\frac{\delta U^2}{P^3},\frac{T}{P^2}),
\label{eqn:under-E}
\end{equation}
by using (\ref{eqn:Langevin-U}).
To first order in $|\dot\phi|^{-1}$, 
we can replace the time average of $\delta U^2$ by the cycle average 
$\overline{\delta U^2}$:
\[
\frac{1}{\tau}\int_{t}^{t+\tau}\delta U(\phi(t'))^2 {\rm d}t'
\rightarrow 
\frac{1}{2\pi}\int_0^{2\pi}\delta U(\phi)^2 {\rm d}\phi\equiv \overline{\delta U^2}.
\]
Applying this approximation, we obtain
\begin{equation}
\frac{{\rm d}E}{{\rm d}t}=-\frac{\gamma\overline{\delta U^2}}{P^2}
+\frac{1}{\tau}\int_{t}^{t+\tau}\delta U \xi(t'){\rm d}t'
+\frac{{\rm d}}{{\rm d}t}(\frac{\tilde p_1^2}{2}+\tilde U) 
+O(\frac{\delta U^2}{P^3},\frac{T}{P^2}).
\label{eqn:under-E-2}
\end{equation}

Since $E=\tilde U+\tilde p_1^2/2+\tilde p_2^2/2$ and 
$|\tilde p_2|\gg|\tilde p_1|$  ($\sim \sqrt{T}$),
(\ref{eqn:under-E-2}) can be rewritten in terms of $\tilde{p_2}$:
\begin{equation}
\frac{{\rm d}}{{\rm d}t}(\frac{\tilde p_2^4}{4})=-\gamma\overline{\delta U^2}
+\tilde{p_2}\{\frac{1}{\tau}\int_{t}^{t+\tau}\delta U \xi(t'){\rm d}t'\}
+O(\frac{\delta U^2}{\tilde p_2},T).
\label{eqn:under-p2-2}
\end{equation}
This equation describes the relaxation to thermal equilibrium
for the second element.  To consider only
the long-time ($\gg O(\tau)$) behavior at the current order of approximation,
 it can be written as
\begin{equation}
\frac{{\rm d}}{{\rm d}t}(\frac{p_2^4}{4})=-\gamma\overline{\delta U^2}
+p_2\delta U \xi(t).
\label{eqn:under-p2-3}
\end{equation}

\subsection{Relaxation at T=0}

When $T=0$, $\xi(t)$ vanishes, and $\tilde p_1$ is negligible 
(or, more precisely speaking, it is $O(P^{-3})$).
Then we obtain the following relations from 
(\ref{eqn:phi}) and (\ref{eqn:p1-U})
\[
\dot\phi = -P+\frac{2\delta U}{P},\quad
p_1 = \frac{\delta U}{P}, \quad
p_2 = P-\frac{\delta U}{P}.
\]
The relaxation equation for the total energy $E$, (\ref{eqn:under-E-2}),
is then simplified as
\begin{equation}
\frac{{\rm d}E}{{\rm d}t} = -\frac{\gamma\,\overline{\delta U^2}}{2E}, 
\label{eqn:under-E-0}
\end{equation}
for $\overline U \ll E$ (that is $E\simeq P^2/2$).
The relaxation equation for the second element is now given by
\begin{equation}
\frac{{\rm d}}{{\rm d}t}(\frac{p_2^4}{4}) =  - \gamma\,\overline{\delta U^2}. 
\label{eqn:under-P-0}
\end{equation}

These relaxation equations reveal that a coupled system in partial
contact with a heat bath exhibits non-exponential, slow relaxation  
after excitation to a high-energy state.
Indeed, from (\ref{eqn:under-E-0}) and (\ref{eqn:under-P-0}),
the form of relaxation is obtained as
\begin{eqnarray}
E^2 &=& E_0^2-\gamma\, \overline{\delta U^2}\, t,
\label{eqn:under-E-sol0}\\
{p_2(t)}^4 &=& p_2(0)^4-4\gamma\, \overline{\delta U^2}\, t,
\end{eqnarray}
where $E_0$ and $p_2(0)$ are the initial total energy 
and momentum of the second element at $t=0$, which 
is defined as a time at which the first element 
reaches thermal equilibrium.
Thus, the characteristic time $t_r$ for the relaxation of $p_2$ 
is given by
\begin{equation}
t_r=\frac{p_2(0)^4}{4\gamma\overline{\delta U^2}}
=\frac{E_0^2}{\gamma\overline{\delta U^2}}.
\label{eqn:result-0}
\end{equation}
With an increase in the injected energy (or momentum), the relaxation time 
increases in proportion to $p_2(0)^4$.  This increase is slower than the
$\exp(p_2(0))$ form obtained from the Boltzmann-Jeans conjecture
\cite{Nakagawa_Kaneko}, 
but even here, there is a rapid increase of $t_r$ with the injected energy.
Comparing the characteristic time (\ref{eqn:result-0}) to that of the standard 
dissipation processes exhibited by systems in full contact with a heat bath
(where $t_r\sim \gamma^{-1}$),
it is confirmed that partial contact indeed results in long-time relaxation
behavior.

\subsection{Relaxation at finite temperature}

Using the expression (\ref{eqn:under-p2-3}), we derive a
relaxation equation for finite temperature. 
The solution of (\ref{eqn:under-p2-3}) can be written
\begin{equation}
\frac{p_2(t)^4}{4}=\frac{p_2(0)^4}{4}-\gamma\overline{\delta U^2}t
+\int_0^t p_2 \delta U \xi(t'){\rm d}t'
\label{eqn:under-sol}
\end{equation}
for $t\gg 2\pi |p_2|^{-1}$.
We next take an ensemble average (indicated by $\langle\cdot\rangle$)
of (\ref{eqn:under-sol}), in order to see the averaged behavior 
for the long-time profile of the relaxation.
Although the ensemble average of the third term, 
$ \langle p_2\delta U\xi\rangle$,
cannot necessarily be decomposed into the product of the average of 
each quantity, 
such a decomposition is valid at the current order of approximation, as
discussed in Appendix 1.  Hence we get
\begin{equation}
\frac{\langle p_2(t)^4\rangle}{4} 
\simeq \frac{\langle p_2(0)^4\rangle}{4}-\gamma\,\overline{\delta U^2}\,t
+\int_0^t \langle p_2(t')\delta U(\phi(t'))\rangle\langle\xi(t')\rangle {\rm d}t'
= \frac{\langle p_2(0)^4\rangle}{4}-\gamma\,\overline{\delta U^2}\,t.
\end{equation}
Since $E\simeq p_2^2/2$ for $|p_2|\gg |p_1|$, this result leads to
\begin{equation}
\langle E^2\rangle = \langle E_0^2\rangle-\gamma\, \overline{\delta U^2}\, t.
\label{eqn:under-E-sol-av}
\end{equation}
Thus the relaxations of the mean values of $E^2$ and $p_2^4$ exhibit the same 
forms as these in the case with $T=0$. 
The validity of the approximations used to obtain these forms have been 
confirmed numerically, as discussed in \S 5, where
(\ref{eqn:under-E-sol-av}) to be valid over a sufficiently long interval.

Furthermore, we can approximate the evolution of the variance of $E^2$ 
(or $p_2^4$) around the mean value,
\begin{equation}
\langle(E^2-\langle E^2\rangle)^2\rangle 
\simeq \int_0^t \int_0^t \langle p_2(t')p_2(t'')\delta U(t')\delta U(t'')
\rangle \langle\xi(t')\xi(t'')\rangle {\rm d}t' {\rm d}t''
= 2 \gamma \,T \int_0^t\langle p_2(t')^2\delta U(t')^2\rangle {\rm d}t',
\end{equation}
where we have again used the decomposition discussed in Appendix 1.
If the time $t$ is not too large, this variance can be approximated by
\begin{equation}
\langle(E^2-\langle E^2\rangle)^2\rangle 
\simeq 2\gamma \,T \,\langle p_2(0)^2 \rangle\,\overline{\delta U^2}\,t
\simeq 4 \, \gamma \,T \langle E_0\rangle\,\overline{\delta U^2} \,t.
\label{eqn:under-E-var}
\end{equation}
This form for the increase of the variance has also be
shown to agree with numerical results, as discussed in \S 5.

{}From (\ref{eqn:under-E-sol-av}), we obtain the mean relaxation time 
$\langle t_r\rangle$ as
\begin{equation}
\langle t_r\rangle\simeq \frac{\langle E_0^2\rangle}{\gamma\overline{\delta U^2}}.
\end{equation}
We thus find that the value of the temperature has little effect on the  
nature of the relaxation from a high-energy state, 
as long as the temperature is sufficiently small.

\section{Overdamped Limit}

When the dissipation is sufficiently large, the
inertial force can be neglected.  In this overdamped limit, 
the equations of motion (\ref{eqn:Langevin}) 
can be written
\begin{equation}
\left\{
\begin{array}{rcl}
p_1&=&\displaystyle
-\gamma^{-1}\frac{\partial U}{\partial \theta_1}+\gamma^{-1}\xi(t)+O(\gamma^{-2}),
\\\\
\dot p_2&=&\displaystyle
-\frac{\partial U}{\partial \theta_2}.
\end{array}
\right.
\label{eqn:overeq}
\end{equation}
Then, for sufficiently large $\gamma$, 
the equation for the energy dissipation is obtained straightforwardly as
\begin{equation}
\dot E = -\gamma p_1^2+p_1\xi(t)
=-\gamma^{-1}{U'^2}+\gamma^{-1}U'\xi(t)+O(\gamma^{-2}),
\label{eqn:over-E}
\end{equation}
where $U'\equiv{\rm d} U/{\rm d}\phi={\partial U}/{\partial \theta_1}
=-{\partial U}/{\partial \theta_2}$.
Solving this equation, we have
\begin{equation}
E=E_0-\gamma^{-1}\int_0^t{U'}^2{\rm d}t'+\gamma^{-1}\int_0^t U'\xi(t'){\rm d}t',
\label{eqn:over-E-sol}
\end{equation}
with the initial energy $E_0$ at $t=0$.

We now consider the case $\gamma \gg P/(2\pi)$.  In other words, 
that in which the time scale of dissipation from
the first element is much faster than the rapid rotation of the second element.
Here, we are interested in relaxation from a high-energy state 
(with very large $|p_2|$) taking place over
a much longer time scale than the rotation time (${2\pi}P^{-1}$).
In this case, the integration over time can be approximated 
by the cycle average:
$
\frac{1}{t}\int_0^t{U'}^2{\rm d}t'\simeq \frac{1}{2\pi}\int_0^{2\pi}{U'}^2{\rm d}\phi
\equiv\overline{U'^2}.
$
By using this approximation, (\ref{eqn:over-E-sol}) can be replaced by
\begin{equation}
E=E_0-\gamma^{-1}\overline{{U'}^2}t+\gamma^{-1}\int_0^t U'\xi(t'){\rm d}t'.
\label{eqn:over-E-sol2}
\end{equation}
An intuitive explanation for this approximation is as follows:
In the limit $P \rightarrow \infty$,  
the change of $\phi$ is expected to be uniform in time.
Then relaxation on the slower time scale can be approximated by
using the uniform average of $U'(\phi)$ over $\phi$, 
to lowest order in $\gamma^{-1}$ and $P^{-1}$.

\subsection{Relaxation at T=0}

For $T=0$, there is no random force, and the relaxation is 
obtained directly as
\begin{equation}
E=E_0-\gamma^{-1}\overline{{U'}^2}\,t.
\label{eqn:over-E-sol0}
\end{equation}
Since $E\simeq p_2^2/2$, this leads to relaxation of the second element 
given by
\begin{equation}
p_2(t)^2=p_2(0)^2-2\gamma^{-1}\overline{{U'}^2}\,t,
\label{eqn:over-p2-0}
\end{equation}
in which $p_2(0)$ is the initial momentum.

{}From (\ref{eqn:over-p2-0}), the characteristic time $t_r$ 
for the relaxation is obtained as
\begin{equation}
t_r=\frac{\gamma p_2(0)^2}{2\,\overline{U'^2}}
=\frac{\gamma E_0}{\overline{U'^2}}.
\end{equation}
Thus in the present case, the dependence of $t_r$ on $p_2(0)$ is different 
from that in the underdamped case, where $t_r\sim p_2(0)^4$.
We remark that even in this overdamped case, the relaxation is much slower than
in the case of full contact with the heat bath ($t_r \sim \gamma^{-1}$).
The crossover between the two limiting cases is discussed on the basis of 
numerical computations in \S 6.

\subsection{Relaxation at finite temperature}

Taking an ensemble average of (\ref{eqn:over-E-sol2}),
we consider the relaxation of the mean value of $E$:
\begin{equation}
\langle E\rangle 
\simeq \langle E_0\rangle-\gamma^{-1}\overline{{U'}^2}t
+\gamma^{-1}\int_0^t \langle U'\rangle\langle\xi(t')\rangle {\rm d}t'
= \langle E_0\rangle-\gamma^{-1}\overline{{U'}^2}t.
\label{eqn:over-E-sol-av}
\end{equation}
Here $\langle U'(\phi)\xi(t')\rangle$ has been decomposed as
$\langle U'(\phi)\rangle\langle\xi(t')\rangle$,
as is justified at the present order of approximation.  
This relation can be intuitively explained as follows.
The quantity $U'(\phi)$ changes due to the Brownian  motion of $\phi$ 
in a manner that is independent of $E$,
because the dependence of $U'(\phi)$ on $E$ appears only at a higher 
order of $E^{-1}$.
Equation (\ref{eqn:over-E-sol-av}) leads to an equation for the relaxation of 
the second element:
\begin{equation}
\langle p_2(t)^2\rangle \simeq 
\langle p_2(0)^2\rangle -2\gamma^{-1}\overline{{U'}^2}\,t.
\end{equation}
It is noted that this form is independent of the temperature of the heat bath.

Fluctuations around the average relaxation path, which
depend on the temperature, can be
straightforwardly estimated from (\ref{eqn:over-E-sol2})
as follows:
\begin{equation}
\langle(E-\langle E\rangle)^2\rangle 
\simeq \gamma^{-2}\int_0^t\int_0^t {\rm d}t'{\rm d}t''\langle U'(t')U'(t'')\rangle\langle\xi(t')\xi(t'')\rangle
\simeq 2\gamma^{-1}T\,\overline{{U'}^2}\,t.
\label{eqn:over-E-var}
\end{equation}
Thus, the distribution of $E$ becomes broader with time,
and this variance is proportional to the temperature $T$ of the heat bath.
The results represented by (\ref{eqn:over-E-sol-av}) and (\ref{eqn:over-E-var})
agree well with the numerical results given in \S 5.

The ensemble average for the relaxation time $t_r$ is obtained 
from eq.(\ref{eqn:over-E-sol-av})
\begin{equation}
\langle t_r\rangle \simeq \frac{\gamma \langle E_0\rangle}{\overline{U'^2}}.
\end{equation}
This is again independent of the temperature.
Thus the discussion given in \S 4.1 is also valid at finite temperature.

\section{Numerical Experiments}

In this section, we check the validity of our analytical approximations,
using results obtained by numerically integrating
the Langevin equation (\ref{eqn:Langevin}) directly.
In order to demonstrate the generality of our results, we adopted 
three forms of the potential $U$ with $0\le U \le 1$ ($|\delta U|\simeq 1$):
\begin{eqnarray}
U_1(\theta) &=& \frac{1}{2}(1-\cos\theta), \nonumber\\
U_2(\theta) &=& \frac{1}{4}\{2 -(\cos\theta + \cos g\theta)\},\nonumber\\
U_3(\theta) &=& \frac{1}{\cosh(\alpha \cos\theta)}-\frac{1}{\cosh(\alpha)}.
\nonumber
\end{eqnarray}
Here, $g$ is the golden mean $(1+\sqrt{5})/2$, and the parameter 
$\alpha$ represents the steepness of the potential $U_3(\theta)$.
$U_1(\theta)$ and $U_3(\theta)$ are periodic functions with period $2\pi$,
while $U_2(\theta)$ is quasi-periodic with two incommensurate frequencies.
$U_3(\theta)$ possesses a steeper form as $\alpha$ increases,
and it approaches a train of $\delta$-functions with period $2\pi$
as $\alpha\rightarrow \infty$.
We set $\alpha=10$ to realize a sufficiently steep shape of the potential.
The values of $\overline {\delta U^2}$ and $\overline {U'^2}$ are
given in Table I for each potential form.

The relaxation at a finite temperature $T$ was studied
by taking an ensemble average.
We integrated (\ref{eqn:Langevin}) 
for almost $1000$ samples with random seeds for each set of parameters,
and thereby obtained a distribution $\rho_t(x)$ of the variable $x(t)$, 
and thus an ensemble average $\langle x(t) \rangle$ 
over this distribution.
For every sample in a given ensemble, we used the same initial conditions 
(given below),
and thus the differences among paths for individual samples resulted 
only from the effect of the random force.

In this section, we study the scaling form for the relaxation of 
the total energy $E$ (not $p_2$) by considering variations of 
$\gamma$, $U(\theta)$ and  $T$.
For this purpose, we chose an initial conditions at a given value of $T$
with initial energy $E_0$ as
\begin{equation}
\theta_1(0)=\theta_2(0)=p_1(0)=0,\,\,
p_2(0)=\sqrt{2E_0-T}.
\label{eqn:initial}
\end{equation}
Here, $E_0$ is a constant value, independent of $T$.
By using these initial conditions, the total energy $E$ was evaluated 
with the initial energy $E_0$ for a number of temperatures,
after a time to allow for the relaxation of the first element.
After this relaxation, it is obtained that $\tilde p_1^2/2\simeq T/2$ and
$\tilde p_2^2/2\simeq E_0-T/2-\overline{U}$ 
(i.e., $E=\tilde p_1^2/2+\tilde p_2^2/2+\overline{U}\simeq E_0$).

\subsection{Underdamped case}

We first give the numerical results for $T=0$.
Figure \ref{fig:under-0} displays the energy relaxation for the 
three kinds of potentials described in Table I.
For the underdamped case, the values $\gamma_u$ 
 in Table I were used for each potential.
The relaxation profile obtained from the numerical experiments is fit well by
$E^2 = E_0^2-\gamma \,\overline{\delta U^2} \,t$, 
for ${\delta U^2}$ corresponding to  each potential form.
Thus the numerical results confirm the analytical expression 
(\ref{eqn:under-E-sol0}). 

As a next step, we examine the relaxation process at finite temperature.
In Fig.\ref{fig:under-av}(a), ensemble averages $\langle E^2\rangle$ are 
displayed as functions of time $t$ during the relaxation process.
This figure includes numerical results from six different set-ups,
using the two potential forms $U_1$ and $U_3$ each with three 
different temperature $T=0.01, 0.1$ and $0.5$.
For a small time $t$, the means $\langle E^2\rangle$ exhibit 
the same scaling relation as in the $T=0$ case:
$\langle E^2 \rangle \simeq E_0^2
-\gamma \,\overline{\delta U^2} \,t$.
This is consistent with (\ref{eqn:under-E-sol-av}).
As the time increases, the mean begins to deviate slightly from the line 
defined by (\ref{eqn:under-E-sol-av}).
At higher temperature, there is a tendency to faster relaxation.

In these computations, the variance of $E^2$ (i.e.
$\langle (E^2-\langle E^2\rangle)^2 \rangle$)
was found to increase linearly with time, as is seen 
in Fig.\ref{fig:under-av}(b).
This figure includes the variances for
the above mentioned six cases, as well as those for two more cases
with a different value of $E_0$ at $T=0.01$ for $U_1$ and $U_3$.
The last two sets were included to examine the effect of changing the
energy $E_0$.
{}From these results, the variance is  shown to satisfy the 
scaling relation
$\langle (E^2-\langle E^2\rangle)^2 \rangle/E_0\, T \simeq 
4\gamma \, \overline{\delta U^2}\,t$,
as expected from the analytical result
in \S 5 (\ref{eqn:under-E-var}).  
Note the slight deviation upward at large time.

\subsection{Overdamped case}

In this subsection, we report the numerical results for the relaxational 
behavior in the overdamped case.
Here the parameter values $\gamma_o$ in Table I were used.
When $T=0$, we found that $E(t)$ beginning from a high-energy state
relaxes as in Fig.\ref{fig:over-0}.
By considering a variety of potential forms (see Fig.\ref{fig:over-0}), 
the result 
$E \simeq E_0-\gamma^{-1} \,\overline{U'^2} \,t$ 
is obtained, confirming the analytical result, (\ref{eqn:over-E-sol0}).

For finite temperature $T$, the ensemble average was studied numerically,
and the same relaxation form,
$\langle E\rangle \simeq E_0-\gamma^{-1} \,\overline{U'^2} \,t$,
is found to hold within the range of temperatures considered, as
displayed in Fig.\ref{fig:over-av}(a).
This figure includes results for six parameter sets:
$U_1$ and $U_3$ each with $T=0.01, 0.1, 0.5$.

The variance of an ensemble exhibits a linear increase with time,
as seen in Fig.\ref{fig:over-av}(b).
{}From the numerical results, it obeys the scaling form
$\langle (E-\langle E\rangle)^2 \rangle/T \simeq 2 \, 
\gamma^{-1} \, \overline{U'^2}\,t$,
which agrees with the analytical relation (\ref{eqn:over-E-var}).
We see that there is slightly larger deviation from the expected scaling 
relation here compared to the 
small deviation exhibited by the mean value in Fig.\ref{fig:over-av}(a).

\section{Crossover Regime}

In this section we study the dissipation of energy in intermediate values
 of damping between
the overdamped and underdamped limits.
In the preceding sections, the relaxation of the
ensemble average of $E$ (or $E^2$) was shown to be independent of
the temperature.
For this reason, we confine our study of the crossover regime 
to the case $T=0$.
The result we obtain here is expected to hold at a finite temperature also.

Recall the relaxational behavior at $T=0$ obtained 
for the underdamped and overdamped limit:
\begin{equation}
\frac{{\rm d}E}{{\rm d}t} =
\left\{
\begin{array}{ll}
- \displaystyle{\frac{ \gamma \,\overline {\delta U^2}}{2}} E^{-1}, \; & P\gg\gamma,\\\\
- \gamma^{-1} \overline {U'^2}, \; & P\ll\gamma.
\end{array}
\right.
\label{eqn:dissipation}
\end{equation}
Here $E\simeq P^2/2$ and $\overline U \ll E$.
In Appendix 2, we obtain more general approximations of $p_1$ and $\dot E$,
for any values of $\gamma$ and $P$.
Here we give an intuitive description of the change in
the relaxational behavior with the decrease of $P$ for a fixed value
 of $\gamma$.

The first element is driven by the second element
and damped by the frictional effect of the heat bath.
When $\gamma$ is sufficiently small, and $P$ is sufficiently larger
 than $O(\gamma)$, this damping due to friction is negligible,
and the dynamics of the first element are approximately 
determined solely by its interaction with the second element.
This leads to $p_1\simeq\delta U/P$.
From this we obtain the upper form in (\ref{eqn:dissipation}).
As the momentum $P$ decreases,
the interaction between the first and the second element becomes 
stronger, and this is accompanied by the growth of $p_1$.
Due to this input of energy to the first element, it rapidly
dissipates energy to the heat bath.
In such a situation, the value of $p_1$ cannot be estimated as $\delta U/P$,
due to the large friction.
Instead, it is estimated as $\gamma^{-1}U'$ for sufficiently 
small values of $P$.
The relaxation equation in this case thus assumes the lower form in
(\ref{eqn:dissipation}).

{}From the above picture, we expect a smooth interpolation between 
the two relaxation forms in (\ref{eqn:dissipation}).
The effect of the heat bath on the first element
is comparable to that of the second element when
the energy $E$ ($\simeq P^2/2$) satisfies the condition
$\gamma\overline{\delta U^2}E^{-1}/2 \sim \gamma^{-1}\overline{U'^2}$.
Taking this into consideration, we define the crossover energy 
for the relaxation behavior,
\begin{equation}
E_c \equiv \frac{\gamma^2 \,\overline{\delta U^2}}{2\, \overline{U'^2}}.
\label{eqn:crossover}
\end{equation}
This represents the energy at which the two relaxational forms are equal.
By introducing $t_c$ defined by
\begin{equation}
t_c \equiv \frac{\gamma^3\,\overline{\delta U^2}}{2\overline{U'^2}^2},
\label{eqn:crossover-t}
\end{equation}
the normalized energy $\varepsilon \equiv E/E_c$, and
the normalized time $\tau \equiv t/t_c$,
(\ref{eqn:dissipation}) can be written as
\begin{equation}
\frac{{\rm d}\varepsilon}{{\rm d}\tau} =
\left\{
\begin{array}{ll}
- 2 \varepsilon^{-1}, \; & \varepsilon\gg 1,\\\\
- 1, \; & \varepsilon\ll 1.
\end{array}
\right.
\label{eqn:dissipation2}
\end{equation}
These equations include neither $\gamma$ nor $U$ explicitly.
Hence, the value of ${\rm d}\varepsilon/{\rm d}\tau$ is expected to depend only on the value of $\varepsilon$,
regardless of the value of $\gamma$ and the shape of $U$.

The relaxation form $\varepsilon$ obtained
numerically is displayed in Fig.\ref{fig:crossover} for the case $T=0$.
To obtain this figure, we computed (\ref{eqn:Langevin}) at $T=0$
for various values of $\gamma$
and plotted the relation between $\varepsilon$ and ${\rm d}\varepsilon/{\rm d}\tau$ 
after the first element reached thermal equilibrium.
We see that the value of ${\rm d}\varepsilon/{\rm d}\tau$ becomes proportional 
to $\varepsilon^{-1}$ as $\varepsilon$ increases beyond $1$, 
while ${\rm d}\varepsilon/{\rm d}\tau$ approaches $-1$
 and becomes independent of the value of $\varepsilon$ 
for smaller $\varepsilon$, as expressed by (\ref{eqn:dissipation2}).
This figure  shows that the relaxation form changes
smoothly between the two limiting regimes.
We also remark that the nature of potential form 
$U(\theta)$ seems to affect the sharpness of the crossover near $E=E_c$.

\section{Summary and Discussion}

In this paper, we have studied the relaxational behavior from a
 far-from-equilibrium (highly excited) state for a system 
consisting of internal and external elements in which only 
the external element is in contact with a heat bath.
In spite of the simplicity of the model, our system exhibits a
power-law-type slow relaxation
 when the potential is bounded, i.e., for $|U|<\infty$.
This slow relaxation is observed
 when the inner part of the system is excited to a high-energy state,
with energy much greater than $|\delta U|$.
The slow relaxation of the internal part results from
its indirect contact with the heat bath. 
In spite of the system's contact with the heat bath, guaranteeing its
approach to equilibrium, the observed slow relaxation
is in a strong contrast with the rapid exponential decay 
exhibited by systems in full contact with a heat bath. 
We therefore find that a Hamiltonian system in partial contact with
a heat bath can maintain its own internal dynamics for a much longer time
 than one in full contact with a heat bath.
If a system has an internal part further separated from the heat bath,
the Hamiltonian dynamics of this part can be preserved much longer.

Suppose the momentum of the second element $p_2$ is constant in time.  Then 
the present model becomes
$\ddot\phi+\gamma\dot\phi+2U'(\phi)=-\gamma p_2+\xi(t)$.
The Brownian motion of a particle in a periodic  potential  $U(\phi)$
in thermal equilibrium has been analyzed using
the Fokker-Planck equation for this model \cite{Risken,periodic}
in the contexts of Josephson tunneling junction \cite{Josephson}, 
super-ionic conductors \cite{ionic_conductor}, 
and two-mode lasers \cite{laser}.
In contrast, we are interested in the
autonomous relaxation of $p_2$ from a highly excited state,
through the internal dynamics obeying (\ref{eqn:Langevin}).
This autonomous change of $p_2$, due to the internal dynamics of 
the composite system, brings about slow relaxation characterized by 
power laws,
even though the system is neither subject to special conditions,
such as being placed at a phase transition point, nor exists in a glassy state
involving many degrees of freedom.

While our interest in Hamiltonian systems in partial contact
with a heat bath originates in our desire to gain an understanding of 
the long-term maintenance of high-energy states,
such systems have also been studied in relationship with 
thermal conduction \cite{FPU}, 
for example, in the situation that two ends of a chain are in contact 
with two heat baths at different temperatures.
According to several simulations performed on systems of this type,
the Fourier law of thermal conduction is not always obeyed.
It appears that certain conditions concerning the internal dynamics, 
as well as a suitable relation of the system with the heat baths, 
are required in addition to ergodicity (or mixing), in order for the
Fourier law to be obeyed.  
These studies also suggest that parts of systems in indirect contact
with heat baths can maintain over extended periods of time 
the characteristic dynamics of isolated Hamiltonian systems.

In a class of Hamiltonian systems in the absence of a heat bath,
very slow relaxation to equilibrium was
suggested in the form of the `Boltzmann-Jeans conjecture'
\cite{Jeans}, and such behavior was reported in Refs.
\cite{Galgani_Benettin,Baldan_Benettin}.
Long-term storage of a large amount of energy within a degree of freedom
through this mechanism has also been studied in a Hamiltonian system
 with chaotic behavior \cite{Nakagawa_Kaneko},
in which localized energy is transferred
 to other degrees of freedom only after a very long time. 
In that case, the energy dissipates logarithmically in time.  
We have reported in this paper that situations exist in which localized
energy in a system dissipates slowly, according to power laws,
 and that long-term energy storage is possible
 even if the system is integrable and in contact with a heat bath.  
{}From this point of view, our result can be regarded as an extension 
of the Boltzmann-Jeans conjecture to include dissipation,
while the relaxational behavior itself is modified from the logarithmic form
of the Boltzmann-Jeans conjecture.
As mentioned in the Introduction, long-term energy storage has been reported 
in some protein molecules \cite{Ishijima_Yanagida},
and the relevance of this storage to molecular motors (for muscle contraction)
has been discussed.  
To apply the present ideas regarding
slow relaxation to proteins, we need to examine relaxation 
from a high-energy states for a system with more degrees of freedom.
This problem will be studied in a future publication \cite{Nakagawa_Kaneko2}.
The main question to be addressed is whether the relaxational behavior 
exhibits a power-law-form or a logarithmic form
when Hamiltonian systems more complex than that considered in the present study
are in partial contact with a heat bath.

\vskip 0.3cm 

This work is supported by
Grants-in-Aid for Scientific Research from
the Ministry of Education, Science and Culture of Japan
(11CE2006, 11837004 and 11740218), and the REIMEI Research Resources 
of Japan Atomic Energy Research Institute.

\section{Appendix 1}

In this appendix, an expression for the ensemble average $\langle p_2^4\rangle$
is derived from (\ref{eqn:under-p2-3}).
Since this equation is a Langevin equation with multiplicative noise,
the drift term of the corresponding Fokker-Planck equation 
 is not necessarily $-\gamma\overline{\delta U^2}$.
We carry out the Kramers-Moyal expansion \cite{Risken} 
for the evolution of the variable $z \equiv p_2^4$
to obtain the Fokker-Planck equation for the probability $P(z,t)$ 
in the form $ \partial P(z,t) / \partial t =  \partial D^{(1)} P 
/ \partial z + (1/2) \partial  / \partial z D^{(2)}   \partial  
/ \partial z P$.
Computing straightforwardly from (\ref{eqn:under-p2-3}),
we obtain the drift term
\[
D^{(1)} = -\gamma\overline{\delta U^2}
+\frac{\delta U^2\gamma T}{\sqrt{2} p_2^2}+o(\frac{T}{p_2^2})
\]
and the diffusion term 
\[
D^{(2)} = p_2^2\delta U^2 \gamma T+o(\frac{T}{p_2^2}),
\]
where the second term of $D^{(1)}$ comes from the multiplicative noise.

Since we confine our study to the situation with sufficiently large $p_2$ 
and sufficiently small $T$, this second term of $D^{(1)}$ can be ignored.
Thus we have $D^{(1)} = -\gamma\overline{\delta U^2}$
to the order of approximation of the current analysis.
Hence, the ensemble average $\langle p_2\delta U\xi\rangle$ can be decomposed
as $\langle p_2\delta U\rangle\langle \xi\rangle$.

\section{Appendix 2}

In this appendix, we give another derivation of the relaxation equation
and a more precise expression of the crossover condition at $T=0$.
Solving the equation of motion for the first element
( $ \dot p_1= -{{\rm d}U(\phi(t))}/{{\rm d}\phi}-\gamma p_1$),
we obtain
\begin{equation}
p_1(t)= -\int_0^{\infty}{\rm d}\tau \,\frac{{\rm d}U(\phi(t-\tau))}{{\rm d}\phi} {\rm e}^{-\gamma\tau}.
\label{eqn:sol1-appendix}
\end{equation}
We write the periodic potential $U(\phi)$ as a Fourier series:
\begin{equation}
\left\{
\begin{array}{lll}
U &=& \displaystyle{\sum_j a_j\cos(\omega_j(\phi-\phi_j))}, \\
U' &=& -\displaystyle{\sum_j a_j\omega_j \sin(\omega_j(\phi-\phi_j))},
\end{array}
\right.
\label{eqn:Fourier}
\end{equation}
where we assume $\overline U=0$ without loss of generality.
Then, assuming that $\phi(t)=\dot\phi\, t$, 
(\ref{eqn:sol1-appendix}) can be written,
\begin{equation}
p_1 = -\sum_ja_j\omega_j\int_0^{\infty}{\rm d}\tau\sin(\omega_j\dot\phi(t-\tau)+\phi_j)
{\rm e}^{-\gamma\tau}
= \sum_j\frac{\dot\phi\omega_j^2 {U_j}
+\gamma U'_j}{(\dot\phi\omega_j)^2+\gamma^2}.
\end{equation}
Here $U_j$ and $U_j'$ are the $j$th components of the expressions in 
(\ref{eqn:Fourier}).
This solution implies that
\[
p_1 = {\dot\phi}^{-1} U, \quad
\dot E = -\gamma\, {\dot\phi^{-2}} {\overline{U^2}} \qquad
(\dot\phi\omega_j \gg \gamma \;\mbox{for all}\; j),
\]
and 
\[
p_1 = {\gamma}^{-1} U, \quad
\dot E = -{\gamma}^{-1}{\overline{U'^2}}
\qquad
(\dot\phi\omega_j \ll \gamma \;\mbox{for all}\; j).
\]
These equations are equivalent to those obtained in \S 6,
but in the presents derivation, the crossover condition is more
explicitly represented as $\dot\phi\omega_j\sim\gamma$.
Since $\overline{U_j^2}=\omega_j^2\overline{U_j'^2}$,
this crossover condition can be expressed as 
\[
\dot\phi^2 \sim \gamma^2 \frac{\overline{U_j^2}}{\overline{U_j'^2}},
\]
which is consistent with the condition discussed in \S 6.

\eject

\begin{table}
\caption{Characteristic values $\overline{\delta U^2}$ and $\overline{U'^2}$
for the three potentials.
$\gamma_u$ and $\gamma_o$ are the values of the 
relaxation coefficient $\gamma$ used in the numerical computations
for the underdamped and the overdamped limit.
}
\end{table}

\begin{figure}
\caption{Energy dissipation for $T=0$.
Relaxation from the initial conditions (\protect\ref{eqn:initial}) 
with $E_0=10$ were computed.  
$E^2$ is plotted versus the scaled time $\gamma\, \overline{\delta U^2}\, t$.
The values of $\gamma$ are those given for $\gamma_u$ in Table I 
for each potential.
With the scaled time $\gamma\, \overline{\delta U^2}\, t$,  
the relaxation curves for the three different potentials coincide. 
This is consistent with the analytical estimate.
}
\label{fig:under-0}
\end{figure}

\begin{figure}
\caption{(a) Relaxation process of the ensemble average 
$\langle E^2 \rangle$ for $T=0.01, 0.1$ and $0.5$, for each of 
the potentials $U_1$ and $U_3$. The  values  of
$\gamma$ are those of $\gamma_u$ given in Table I, while
the initial conditions are given by (\protect\ref{eqn:initial}) with $E_0=10$.
The scaled time $\gamma\,\overline{\delta U^2}\, t$ is used for each potential.
The line indicates the analytical approximation represented by 
(\protect\ref{eqn:under-E-sol-av}).
(b) Temporal evolution of the variance of $E^2$ corresponding to the
six cases in (a)
and also for two additional cases with  $E_0=12$ and $T=0.01$ 
for $U_1$ and $U_3$.
The line indicates the analytical approximation represented by
(\protect\ref{eqn:under-E-var}).
}
\label{fig:under-av}
\end{figure}

\begin{figure}
\caption{Energy dissipation for $T=0$.
The value of $\gamma$ fpr each potential is the corresponding value of 
$\gamma_o$ given in Table I,
and the initial conditions are given by (\protect\ref{eqn:initial}) 
with $E_0=10$.
With the scaled time $\gamma^{-1}\, \overline{U'^2}\, t$,
the relaxation curves for the three different potential coincide, in
agreement with the analytical approximation.
}
\label{fig:over-0}
\end{figure}

\begin{figure}
\caption{(a) Relaxation process of the ensemble average 
$\langle E \rangle$ for $T=0.01, 0.1$ and $0.5$, starting from 
the initial conditions (\protect\ref{eqn:initial}) with $E_0=10$.
The potential forms used are $U_1$ and $U_3$, while the parameter values
$\gamma$ are given by  $\gamma_o$ in Table I, corresponding to the 
overdamped case.
The scaled time $\gamma^{-1}\,\overline{U'^2}\, t$ is for each potential. 
The line indicates the analytical approximation represented by
(\protect\ref{eqn:over-E-sol-av}).
(b) Temporal evolution of the variance of $E$ corresponding to
the six cases in (a).
The line indicates the analytical approximation represented by
(\protect\ref{eqn:over-E-var}).
}
\label{fig:over-av}
\end{figure}

\begin{figure}
\caption{Dependence of $\dot E$ on $E$ 
for a variety of values of $\gamma$, 
which range from the underdamped to overdamped case.
Here $E_c$ and $t_c$ are determined for 
each potential according to the value of $\gamma$
(see (\protect\ref{eqn:crossover}) and (\protect\ref{eqn:crossover-t})).
The functional forms of $\dot E(E)$ for the two potentials 
$U_1$ and $U_3$ are essentially identical, 
except in the intermediate range around $E/E_c=1$.
}
\label{fig:crossover}
\end{figure}


\begin{thebibliography}{999}
\bibitem{note}
One might think that for any part of a protein molecule,
the remaining parts play a role of a heat bath, since the molecule
has always large degrees of freedom.
This is not always the case, or even if it is true, the `heat bath' there
cannot be regarded as a random process without any structure.
It is known that a class of Hamiltonian systems with large degrees
of freedom exhibits spatially or temporally correlated motion.
Indeed, there appears some order such as clustered motion
\cite{Konishi_Kaneko},
or some part, highly excited, can be isolated from the other part
with very slow relaxation \cite{Jeans,Galgani_Benettin,Baldan_Benettin}.
\end{thebibliography}
\end{document}